\documentclass[12pt]{iopart}
\usepackage{graphicx}

\begin{document}

\title[Integrated Nuclear Simulation Data Analysis]{A domain-specific analysis
system for examining nuclear reactor simulation data for light-water and sodium-cooled fast reactors}
\author{Jay Jay Billings$^1$\footnote{Corresponding author. Telephone:
+1 865 241 6308, Email: billingsjj@ornl.gov, Twitter: @jayjaybillings}, Jordan
H Deyton$^1$,S. Forest Hull III$^1$\footnote{Present address: Florida Institute
of Technology, 150 W. University Blvd. Melbourne, FL, 32901, USA}, Eric J
Lingerfelt$^1$ and Anna Wojtowicz $^1$}
\address{$^1$ Oak Ridge National Laboratory PO Box 2008 MS6173 Oak Ridge,
TN,37831, USA}

\begin{abstract}
Building a new generation of fission reactors in the United States presents many
technical and regulatory challenges. One important challenge is
the need to share and present results from new high-fidelity, high-performance
simulations in an easily usable way. Since modern multiscale, multi-physics
simulations can generate petabytes of data, they will require the development of
new techniques and methods to reduce the data to familiar quantities of interest
(e.g., pin powers, temperatures) with a more reasonable resolution and size.
Furthermore, some of the results from these simulations may be new quantities
for which visualization and analysis techniques are not immediately available
in the community and need to be developed.

This paper describes a new system for managing high-performance simulation
results in a domain-specific way that naturally exposes quantities of interest
for light water and sodium-cooled fast reactors. It describes requirements to
build such a system and the technical challenges faced in its development at all
levels (simulation, user interface, etc.). An example comparing results from two
different simulation suites for a single assembly in a light-water reactor is
presented, along with a detailed discussion of the system's requirements and design.
\end{abstract}


\section{Introduction}

Modeling and simulation has always played a vital role in nuclear engineering
and its applications. It is becoming even more relevant as the community starts
examining and constructing new types of reactors. Although existing
computational models of the conventional fleet perform very well within the
operational experience base of today's power plants, they do not necessarily
provide the predictive capability needed to enable the deployment of completely
new designs or to enable the use of today's designs for very different modes of
operation. New simulation codes are under development in several efforts
sponsored by the US Department of Energy's Office of Nuclear Energy, including
those from the Nuclear Energy Advanced Modeling and Simulation (NEAMS) program
\cite{NEAMS} and the Consortium for Advanced Simulation of Light Water Reactors
(CASL) \cite{CASL}.

These new codes are much different from the existing codes. Whereas
tried-and-true codes employ sophisticated engineering models calibrated based
on experimental observation to describe system behavior, the new code suites
leverage high-performance computing (HPC) platforms to provide truly predictive
simulators. The new codes can examine the physics of nuclear reactors with an
unprecedented resolution at spatial and temporal scales ranging from the
microstructure of the fuel all the way to the plant itself, and they are
commonly capable of coupling neutronics, fuel mechanics, structural
mechanics, and thermohydraulics \cite{MOOSE},\cite{SHARP},\cite{VERA2013}.
While they are very powerful, they inevitably generate many terabytes and even
petabytes of data that greatly exceed what any single person can absorb and
interpret alone. Even results that are refined through post-processing can still
be gigabytes in size.

Additional challenges arise when the user actually examines and reviews the
results. Large tables of data, even if well-organized, are not easily
consumed by humans because of their size and general lack of sufficient context
to eliminate ambiguity. Furthermore, in this particular area, there are often
very large numbers (fluxes, $\approx10^{12}$) associated with very small numbers
(displacements, $\approx10^{5}$) that make intuitive interpretation of the
rsults difficult. More often than not, users are forced to implement
their own codes on top of the post-processing routines to collect, assimilate,
rescale, normalize, and plot the very small amount of data that they actually
need, which is a very costly chore in terms of both time and resources. Comparing the
results from new simulations to those from old simulations or experiments also
requires custom code, thereby multiplying the cost of adoption.

The authors assert that it is not sufficient to leave these challenges
unaddressed and ``pitch it over the fence'' to users and analysts. Instead,
just as new codes are under development to address the physics questions, new
technologies need to be developed to address data analysis. The system
presented in section \ref{sec:arch} seeks to address this issue in
three ways. First, it provides specialized data structures and input/output
(I/O) libraries designed specifically for storing quantities of interest from
nuclear simulations. Second, it provides a user interface that complements the
I/O libraries to provide highly tailored views that put data in the proper
context and reduce ambiguity. Finally, it provides an extension interface for
adding custom analysis routines that can be easily coupled to routines for data
mining and tailored analysis. An example of the utility of the system provided
in section \ref{sec:parts} demonstrates the ease with which interesting
information between two codes, one new and one seasoned, can be extracted from
the system.

\section{Architecture}
\label{sec:arch}

\begin{figure}[h!]
\begin{center}
\caption{A high-level Unified Modeling Language class diagram of the system's logical architecture,
highlighting its focus on interacting with different Parts, or IReactorComponents, of the
reactor data.}
\label{fig:architecture}
\includegraphics[width=\textwidth]{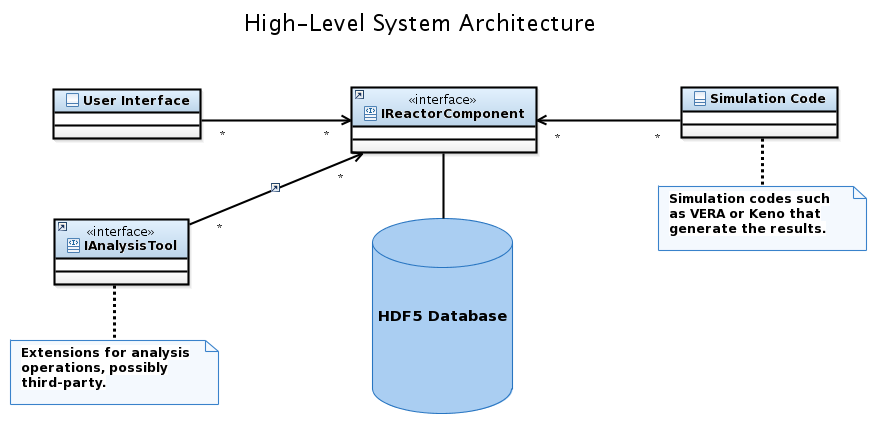}
\end{center}
\end{figure}

A high-level overview of the system's logical architecture is presented in
Fig.\ \ref{fig:architecture}. The system is split into three parts, one each for
users who examine data, developers who store data, and those interested in
implementing custom analysis routines to examine the data. In each case, care
was taken to make sure that the workflow of the respective actor was easy and
intuitive for that use case.

Five high-level functional and nonfunctional requirements were used to guide
the detailed design of the system based on feedback from interviews and experiments
with prototypes, all of which were derived from the assumption that the
simulation results must be shared with one or more people.
\begin{enumerate}
  \item The data should be organized according to the natural layout of a nuclear
reactor and not on the discretized geometry of the simulation code.
  \item A data point has a position in three dimensions plus time with a value,
uncertainty, and units.
  \item A reactor is composed of \textit{Parts}, which are represented by custom
  data structures, and data can be stored for every Part (literally \textit{on} every
  Part) available.
  \item The information is stored in a cross-platform, self-describing, binary
compatible, open format.
  \item The data must be easily and equally accessible from languages used for
  user interfaces \textit{and} from languages used to author simulation codes.
\end{enumerate}

The need to compare data from two simulations led directly to requirement 1.
Storing output data on the grid or mesh used by the simulation would allow only
immediate and direct comparisons with other results from the same code unless
grid-point mapping or mesh-to-mesh transfers \cite{tautges_2009} were performed.
However, storing data based on the layout of a reactor---on cores, assemblies,
pins, etc.---allows for immediate comparison of results, albeit at the cost of
pulling that information from the grid or mesh during the simulation. One drawback 
of this design is that some data \textit{could} be lost when the grid or
mesh is removed. However, this is not strictly the case, since data can be
mapped to any Part in the reactor and given any position.

It is extremely important that the data be stored in a binary compatible
open format available on multiple platforms and readable in
multiple languages (requirements 4 and 5) for three reasons. First, the subject
matter experts will examine the data on workstations, not clusters or
supercomputers. Second, in many situations, it must be stored in such a way that
it can be reviewed if required, possibly at a much later date. Finally, someone
may need to view the data who does not have the expertise to write the code to
read the database but is otherwise qualified to evaluate it via a user
interface.

The definition of a reactor, requirement 3, loosely follows that of real light
water reactors (LWRs) and sodium-cooled fast reactors (SFRs). Reactors are
hierarchically composed of a set of Parts, where a Part describes both the
large-scale structures of a reactor, such as assemblies, and smaller Parts like
pellets or ``material blocks.'' Data is stored for these Parts according to the
description in requirement 2, including time, space, uncertainty and dimensional
information, in addition to the value of interest.

Exact descriptions of the Parts of the nuclear reactors supported by this
project are provided in section \ref{sec:parts}. The Parts are exactly the same
in both supported languages, the user interface, and the I/O libraries.
The design is general and based on publicly available information from
Wikipedia, the Nuclear Regulatory Commission's public website, and interviews
with subject-matter experts. No proprietary designs are considered, and no
proprietary information is used or provided, although it is certainly possible
to add such Parts to the system with minimal effort.

The Parts are organized in a class hierarchy of data structures that uses
object-oriented design principles to take advantage of both the natural
hierarchy of reactors and abstractions for similar Parts (inheritance). The
two different types of reactors that are currently supported---pressurized water
reactors (PWRs) and SFRs---are available in their
own modules but are accessed in the same way.

\subsection{Input-Output Libraries}

\begin{figure}[h!]
\begin{center}
\caption{A Unified Modeling Language sequence diagram that shows the basic process by which the
results computed by a simulation code are loaded in the user interface (``NiCE'')
via the input-output interface, (``NiCE I/O API'').}
\label{fig:sequenceDiagram}
\includegraphics[width=\textwidth]{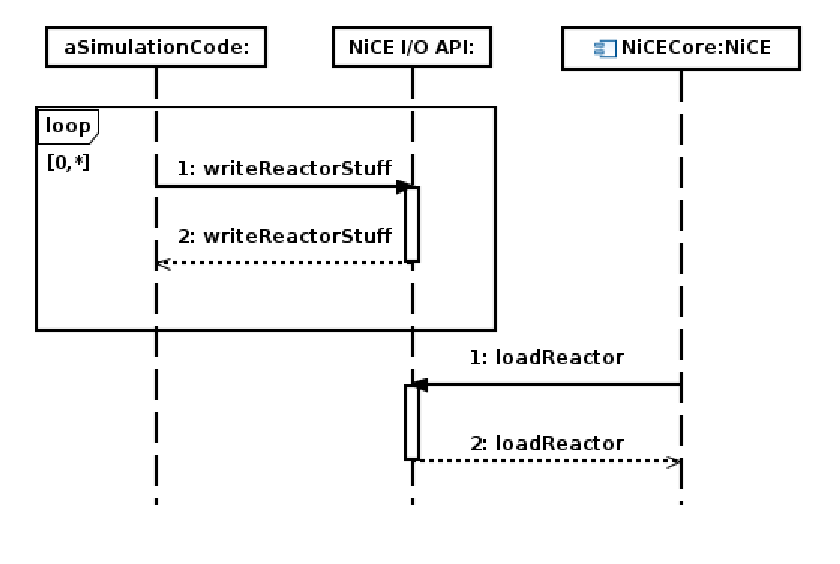}
\end{center}
\end{figure}

The system provides its own libraries for managing input and output to relieve
software developers from the task of writing their own file parsers and
emitters that conform to the shared specification. They remove the burden
of writing to older data formats for developers of new codes and remove
the complication of writing to newer formats, which may require significantly
specialized skills, for maintainers of existing codes. Instead, developers
instantiate and fill the Parts with data and then store those parts using the
provided I/O routines. The workflow is shown, at a high level, in Fig.\
\ref{fig:sequenceDiagram}. In this figure, the ``NiCE I/O API'' represents
the I/O library, and it acts as an intermediary between the simulation code and
the user interface. The process is relatively simple: the developer loads the
data structures needed and tells the library to write them. A user, on the other
side, picks a file and the user interface loads that file.

Developers may store whatever data they need on any Part. Data is stored
\textit{associatively} so that sets of data are stored against tags and any
tag can be used. It is trivial to store data tagged with ``Axial Power''
right next to data tagged as ``Cross sections'' or ``Velocity'' on the same
Part. Writing the data is then very simple: the developer calls a single write
operation on the ``highest'' Part of the reactor,, which is normally the
reactor core, that contains all of the other Parts the developer has. The simulation
code may write multiple reactors to disk, (thus the loop in Fig.\
\ref{fig:sequenceDiagram}). Any client can read this file once it is written,
such as the user interface or standalone analysis codes, using the
inverse read operation with the desired data file passed as an input argument.

The file layout, which is different from the \textit{format}, of the I/O
libraries matches the class hierarchy of the data structures with a few
exceptions. For example, it is much more memory-efficient to store the units of
values in individual tables and refer to the name of the units with an integer
in the file because, in practice, the integer is much smaller than a
fixed-length string for representing an arbitrary system of units. A much more
important optimization is the reuse of large structures like assemblies and
pins, where possible. The layout of a physical reactor will have only a small
number of different pin types (or assemblies) relative to the total number of
pins in the reactor. Exploiting this fact by reusing the geometry and material
information both in the data structures and on disk greatly reduces the amount
of memory required. This detailed analysis of the layout of the data has made
it possible to read and write data for ``full'' cores with 50 axial levels,
hundreds of assemblies, and 50,000 pins in seconds.

\subsection{User Interface}

The ultimate goal of this system is to put simulation data into the hands of
human users, and humans are much more efficient at examining visual images than
raw data. The system includes a graphical user interface that
presents information in multiple ways, including custom views, color maps, plots,
and raw data. The user interface is designed so that users \textit{feel like
they are manipulating a reactor instead of reading a paper}.

The most important design requirement for the user interface is that it enable
users to perform comparative analysis. That is, users must be allowed to
compare results among multiple reactors side-by-side, with at least one of the
reactors accepted as a ``gold-standard'' reference. For that matter, users
must be allowed to examine the same part of a reactor with multiple views of
its own data---geometry, color maps, plots, and so on.

Each Part has at least two associated graphical views, one each for geometric
information, including assigned materials, and state point data, which is defined
as the values of the quantities of interest at a given timestep. Views can be
switched via a simple toolbar or by right-clicking on the canvas. It is
possible to create plots for state point data for some Parts and, using
specialized analysis routines, compute direct numerical comparisons (and create
plots of those comparisons) between the state point data for two or more Parts.
Each view can be saved as a PNG image file with a handy button on the toolbar.

Three-dimensional core geometries are supported by the system, but the user
interface breaks this into two xy and z views. The xy view shows a cross
section of the current Part at the specified axial level. The z view is next to
the xy view and shows the axial geometry.

The user interface also exposes any external analysis routines that are
available to users in a menu, as well as any options that can be configured for
those routines. Results from the analyses are captured and presented to the
user in a simple list. The file names on the local hard drive and their associate
timestamps are also presented so that users can easily move the data to another
system if needed. One external analysis routine is provided by default---to
compute percentage differences among pins---and it is treated as a somewhat special
case in that its results are automatically plotted, which is not the case for
other analyses.

It is possible to view the data stored in the system using other user interfaces
and techniques, c.f. section \ref{subsec:Technologies}.

\subsection{Integration with External Analysis Routines}
\label{subsec:analysisTools}

The system can be extended to use external analysis routines for teams or
developers who want to manipulate the data in a specific way. This mechanism is
provided because it is impossible for the core development team to guess all
of the possible analyses, much less implement and support the capability in the
long term. The routines are referred to as ``external'' simply because they are
plug-and-play.

External routines are exposed to users through the user interface. The external
routines can be anything from simple math operations to large wrappers that ship
the work off to other libraries. The development team has tested three such
extensions to date: the differencing tool for pin powers mentioned previously, a
k-means clustering routine to find clusters in the data, and a set of
``routines'' that wrap a very large visualization toolkit, VisIt \cite{visit},
to create different plots from the default set. The differencing routine is the
only external routine available by default, but more detailed information on 
findings with the k-means clustering routines can be found in
\cite{pokhriyal_ans_2013}. K-means is a clustering algorithm used to find
groups in data that can later be used for classification. Ideally, a set of
external analysis routines would be available to perform classification and
anomaly detection on pins and assemblies as a form of automatic data triage to
locate troublesome, erroneous, or simply interesting areas of the core.

Developers can add routines by implementing the ``IAnalysisTool'' interface
shown in Fig.\ \ref{fig:architecture}, as well as a few other interfaces
associated with it, and declaring the analysis tool as a service to the
framework. Analysis routines are written in Java and are dynamically
consumed by the framework at runtime. Currently, only assemblies are passed to
analysis tools, not reactor cores. This capability is not available in the C++
implementation, because that implementation is primarily focused on I/O for the
simulation codes. Much more detailed descriptions for all of these interfaces are 
presented in the project documentation at the NiCE project page
\cite{NiCEPage}.

\subsection{Technologies}
\label{subsec:Technologies}

The system is built with mostly off-the-shelf components and assembled as part
of the NEAMS Integrated Computational Environment (NiCE),
\cite{NiCEPage}\cite{NiCE_eclipseCon2013}\cite{NiCE_eclipseCon2014}. The
code for developers is written in C++ so that it can be easily used with
C/C++ and Fortran codes. Version five of the Hierarchical Data Format (HDF5)
\cite{hdf5} is used to read and write the data structures to and from disk.
This greatly reduces the amount of work required to satisfy the cross-platform,
multi-language requirements, as HDF5 is natively available in many languages
and is completely open. HDF5 also provides multiple ways to access the data in
``hdf'' files in Ascii, including a command line utility---h5dump---that will
dump the contents to disk and a graphical utility---hdfview---for viewing the
contents of any HDF5 file.

The graphical user interface is implemented in the same system as NiCE, which
uses the Eclipse Rich Client Platform (RCP) \cite{eclipseRCP}, and is written
in Java. RCP is the platform on which the Eclipse Integrated Development
Environment is built, is very flexible, and is cross-platform. It has many
tools and utilities for working with graphics and connecting those graphics to
data.

The k-means algorithm discussed in \ref{subsec:analysisTools} was
implemented in-house for NiCE, based on open literature, and is available in the
source code. It is not enabled by default because the development team is
exploring the possibility of using third-party capabilities to replace the
home-grown version, namely Apache's Mahout \cite{mahout}, and enable more data
mining capabilities.

\section{Parts}
\label{sec:parts}

The system is composed of many Parts across the different levels of the reactor
and is capable of storing data for PWRs and
SFRs. The following discussion presents each Part
in the system, the types of data that can be assigned to the Parts, and the
visualizations available in the user interface. 

The data presented for PWRs is taken from real simulations performed with the
Virtual Environment for Reactor Analysis (VERA) \cite{VERA2013} and KENO
\cite{ornl:scale} for problem 3a of the VERA Benchmarks \cite{godfrey_2013}.
This problem represents a single 17 by 17 PWR fuel assembly at the beginning of
its life and at constant temperature. The assembly is refined with 49
axial levels. VERA was modified for our purposes to write the results directly to
file using the data structures and I/O libraries from our system. We obtained
the KENO results from the author of \cite{godfrey_2013} and converted them into
our system using a stand-alone program. Additional assemblies were added to the
geometric configuration of problem 3a to show the system's ability to work with
assemblies other than fuel assemblies, such as in-core instruments.

The results shown in the pictures for SFRs \textit{are generated and do not
represent a physical system}. The SFR pictures are presented solely to instruct
readers about the capabilities of the data structures and the user interface. 

All of the graphical views below exploit object-oriented design principles to
minimize the amount of custom code needed to draw the different shapes. In many
places, the same code is reused to draw Parts from PWRs and SFRs. The simplicity
of using this system is also noteworthy, as the estimated time for a new user to
go from importing data to exporting a plot like those shown below is merely
minutes and does not require any knowledge of file formats, data layouts, or
scripting languages.

\subsection{Part Properties}

\begin{figure}[h!]
\begin{center}
\caption{A view that shows the properties of the selected part, including its
composition and geometry.}
\label{fig:ReactorProperties}
\includegraphics[width=\textwidth]{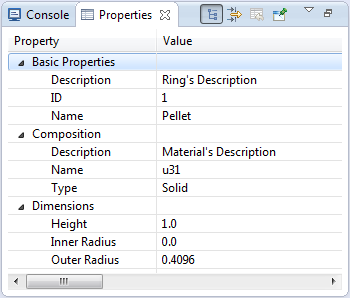}
\end{center}
\end{figure}

Each Part has a set of properties assigned to it that represent the most common
properties of the given piece, such as material, pitch, diameter, and so on. These
properties can be manipulated programmatically and are viewable within the user
interface in a ``Properties View,'' as shown in Fig.\
\ref{fig:ReactorProperties}.

Spatial data about a Part is stored by setting a \textit{data provider} that
includes \textit{data} elements as described by requirement 2 in the Architecture section. Data
providers are simply containers and are capable of managing arrays of data at
different times. Data providers can be stored on any Part (as required by the
IReactorComponent interface). However, in practice, this can destroy performance,
so some Parts manage the data providers for other Parts.

\subsection{Cores}

Reactor cores are the highest Parts in the hierarchy at present. The
PressurizedWaterReactor class represents PWRs. This class extends a generic LWR
base class that defines some properties and operations common across different
types of LWRs, including both PWRs and boiling water reactors\footnote{The
latter is not yet fully supported and therefore is not discussed here.}.
Assemblies of any type are stored in their own grids, so that one type of
assembly can share the same position as another type, which allows for control
rod (``spider'') assemblies to be added. Control banks, fuel assemblies,
in-core instruments, and ``rod cluster'' assemblies can be added to PWRs.
Additionally, the fuel assembly pitch and the grid labels can be configured on
the PressurizedWaterReactor class.

\begin{figure}[h!]
\begin{center}
\caption{A view of a pressurized water reactor core with one fuel assembly
(green) and many control banks (yellow).}
\label{fig:PWRCore}
\includegraphics[width=\textwidth]{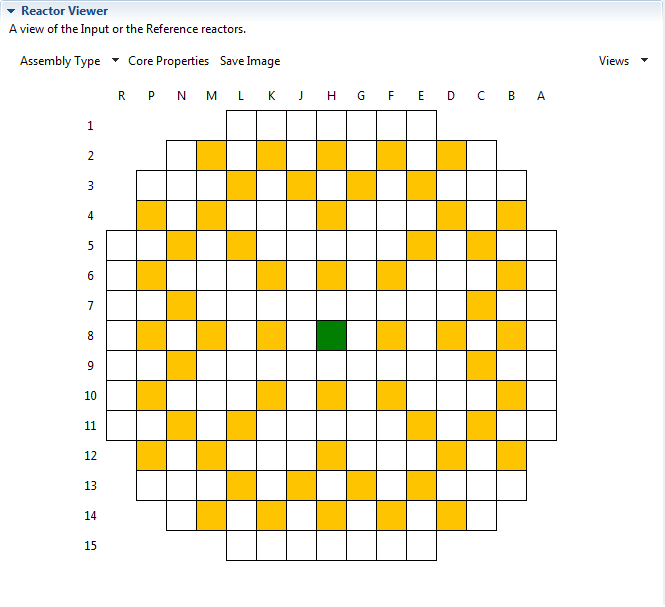}
\end{center}
\end{figure}

Figure\ \ref{fig:PWRCore} shows the geometric view of a PWR core in the user
interface for the 3a problem. The grid labels are read as configured on the
PressurizedWaterReactor class and displayed along the top horizontal and
left vertical axes. Each assembly is represented by a single square on the
grid. The single green assembly in the middle is a 17 by 17 fuel assembly, and the
yellow assemblies are control banks. The type of assembly shown can be
configured using the ``Assembly Type'' dropdown button.

The SFReactor class represents SFRs. Its interface is nearly identical to that of the
PressurizedWaterReactor class, but the allowed assembly types and properties are
different. Fuel, control, reflector, shield, and test assemblies can be added to
SFRs. As for PWRs, size can be obtained for SFRs; but instead of only an
assembly pitch, SFRs offer both a lattice pitch and a flat-to-flat distance.
Flat-to-flat distance is stored on the core for convenience, and the system
assumes that all assemblies have the same flat-to-flat distance.

\begin{figure}[h!]
\begin{center}
\caption{A view of a small sodium-cooled fast reactor core with fuel
assemblies in yellow and green (selected).}
\label{fig:SFRCore}
\includegraphics[width=\textwidth]{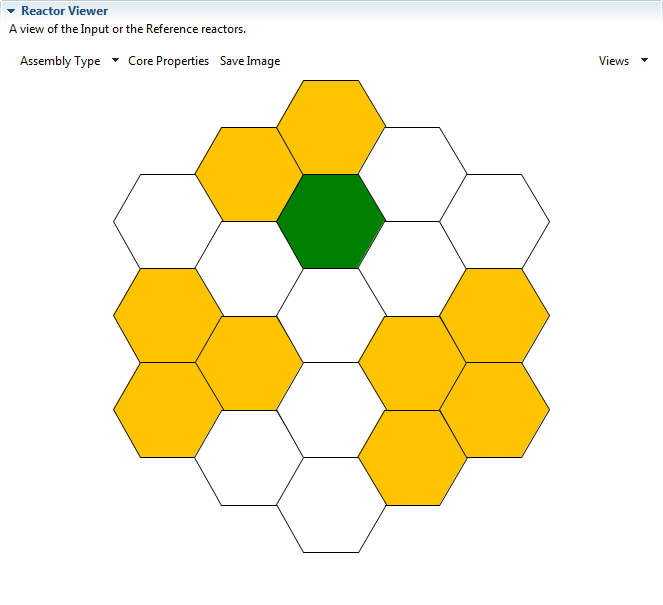}
\end{center}
\end{figure}

The graphical representation of an SFR with fuel assemblies is shown in Fig.\
\ref{fig:SFRCore}. The view shows the hexagonal layout of the core and functions
exactly like the view of the PWR core. 

\subsection{Assemblies}

Multiple assembly types can be configured for both PWRs and SFRs. Each assembly
is composed of a collection of rods, and each rod is mapped to a location on a
grid. ``Rods'' in this case are not necessarily fuel rods. They could be
control rods, for example. The data providers that contain state point data for
the rods, of any type, are stored on the assemblies at a specified grid location
instead of directly on the rods. This is done to optimize data access times and storage
(although they can still be stored on the rods themselves if needed).

PWR assemblies can store their sizes and rod/pin pitches. SFR assemblies can
store their size duct thicknesses in addition to their sizes.

SFRs make a distinction between a ``pin'' and a ``rod,'' which is discussed in
detail in section \ref{subsec:pins}. PWRs make no such distinction and refer to
both simply as ``rods.''

PWRs support
\begin{itemize}
  \item\textbf{Fuel assemblies} of burnable fuel that is used in the
  reactor.
  \item\textbf{Control banks} used to regulate the power within the reactor.
  \item\textbf{Incore instruments} that represent assemblies configured with
  detectors or sensors within a core.
  \item\textbf{Rod cluster assemblies} of (typically neutron absorbing) rods
  placed in and moved between fuel assemblies during refueling outages.
\end{itemize}
 
SFRs support
\begin{itemize}
  \item\textbf{Fuel assemblies} of burnable (inner fuel, outer fuel) and
  blanket (optional) assemblies.
  \item\textbf{Control assemblies} that represent primary and secondary
  (shutdown) assemblies. 
  \item\textbf{Reflector assemblies} that are configured to reflect neutrons.
  \item\textbf{Shield assemblies} for shielding against radiation.
  \item\textbf{Test assemblies} that represent assemblies used for testing
  materials and fuels.
\end{itemize}

\begin{figure}[h!]
\begin{center}
\caption{A geometric representation of one quarter of a pressurized water
reactor fuel assembly with control rods in blue and fuel pins in red.}
\label{fig:PWRAssembly1}
\includegraphics[width=\textwidth]{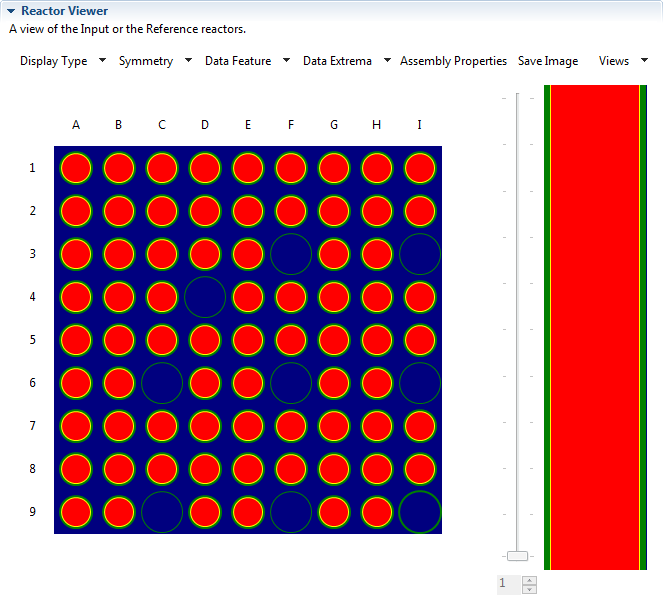}
\end{center}
\end{figure}

The geometry of one quarter of the PWR fuel assembly from problem 3a is shown in
Fig.\ \ref{fig:PWRAssembly1}. Rods are represented by circles, with control rods
represented by blue circles and fuel rods represented by red circles. The blue
spaces between rods represent the coolant. The rods are arranged according to
the pitch and size configured on the assembly. The ``main'' view on the left
represents the axial level (1 in the figure) selected using the slider, spinner,
or clickable axial view on the right. Like the core view, the grid labels are
taken on, configured on, and read from the assembly and displayed on the top
horizontal and left vertical axes.

\begin{figure}[h!]
\begin{center}
\caption{Axial power mapped onto an assembly from a pressurized water reactor
with areas of higher power in red and lower power in blue.}
\label{fig:PWRAssembly2}
\includegraphics[width=\textwidth]{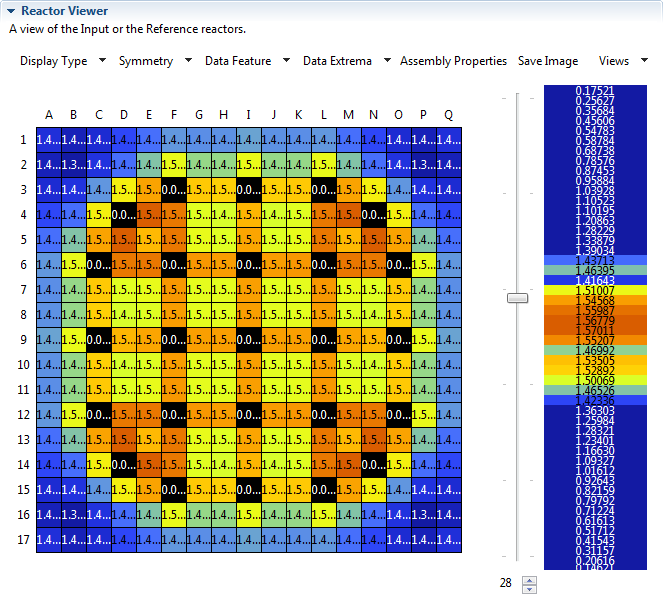}
\end{center}
\end{figure}

The tool bar can be used to switch this view from the geometric configuration to
a view of the simulation data, as depicted in Fig.\ \ref{fig:PWRAssembly2}. The
axial pin powers for problem 3a for the entire fuel assembly are shown in this
view. Each square represents a rod, and the color mapping is bluer for lower
values and redder for higher values. (The circles from the geometric view have
been replaced with squares to make it easier to show the data values.) The
color mapping in this figure was normalized for the selected axial level (level
28), but it can also be set relative to the whole assembly or all assemblies.

\begin{figure}[h!]
\begin{center}
\caption{A geometric representation of a seven-pin sodium-cooled fast reactor
assembly completely composed of fuel rods.}
\label{fig:SFRAssembly1}
\includegraphics[width=\textwidth]{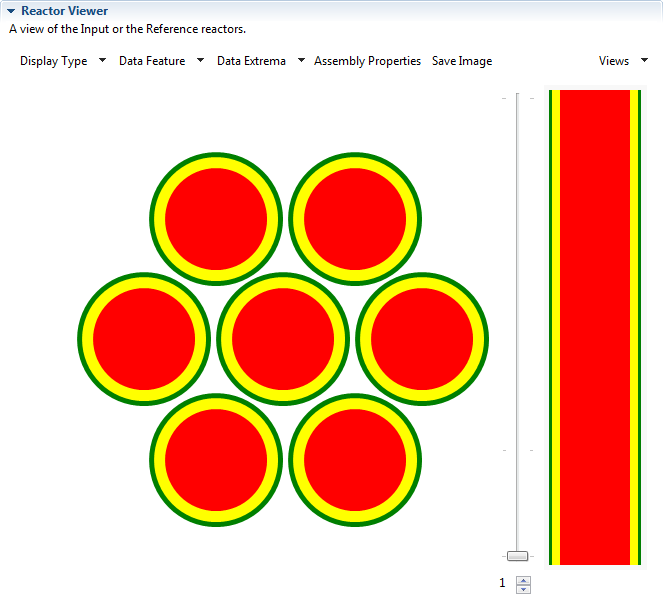}
\end{center}
\end{figure}

\begin{figure}[h!]
\begin{center}
\caption{A view of random data mapped to the seven-pin sodium-cooled fast
reactor assembly to show the color mapping capability.}
\label{fig:SFRAssembly2}
\includegraphics[width=\textwidth]{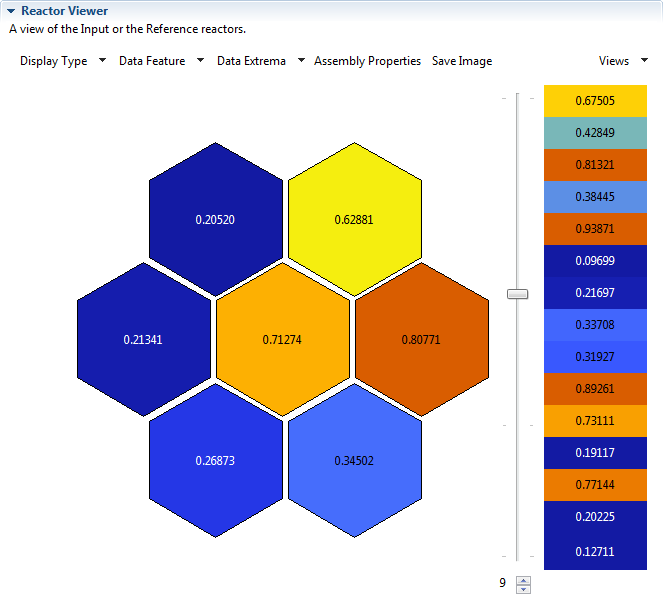}
\end{center}
\end{figure}

Figures\ \ref{fig:SFRAssembly1} and \ref{fig:SFRAssembly2} show the same views for
SFRs. Both views have different grids from their PWR counterparts, but the same
principles apply. Each circle in the geometric view represents a fuel pin in the
SFR assembly and each hexagon in the data view represents the same fuel pin.

The type of data displayed in the data view, for either reactor type,
can be switched with the ``Data Feature'' button. For the PWR example in problem 3a,
both axial and total pin power data are available and can be selected. Any data
stored on the assembly in a data provider will appear in the list and
can be selected.

\subsection{Pins and Rods}
\label{subsec:pins}

PWR assemblies are filled with rods made of either fuel or poisons or, in some
cases, simply empty. The structure of a rod is defined by a collection of
``material blocks'' that describe the materials in a rod between two points
along its axis. Material blocks are axisymmetric and are composed of concentric
rings of materials. For example, a material block in a PWR fuel rod has a fuel
ring, a fill gas ring, and a cladding ring. The clad and fill gas are typically
defined separately from the material blocks for the fuels. Each rod can also
store a pressure. 

Each ring is defined by an inner and outer radius, a height, and a material
type. Materials can be gases, liquids, or solids and can be labeled to match the
materials in the simulation.

The SFR capability makes a further distinction between ``pins'' and ``rods.''
In SFRs, the pin is the basic unit of fuel assemblies and control assemblies.
Rods are used only in reflector assemblies. Pins are identical to rods in
PWRs.

\begin{figure}[h!]
\begin{center}
\caption{A geometric representation of a single fuel pin in a pressurized water
reactor with red, yellow, and green rings for the fuel, fill gas, and cladding,
respectively.}
\label{fig:PWRPin1}
\includegraphics[width=\textwidth]{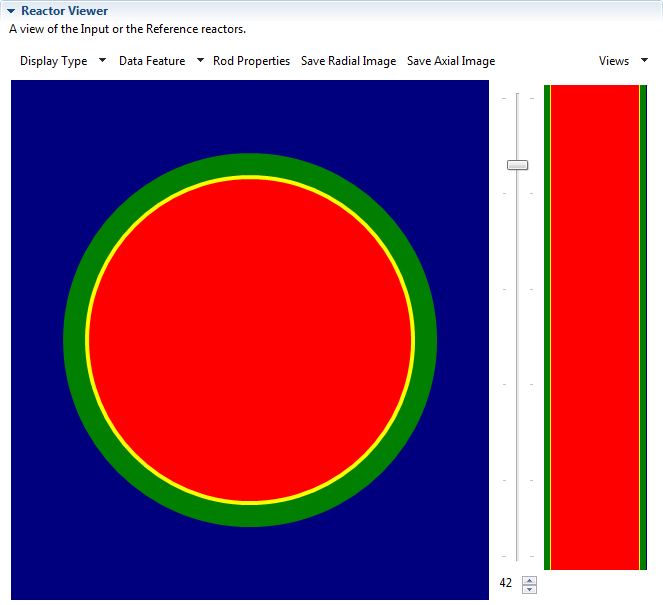}
\end{center}
\end{figure}

\begin{figure}[h!]
\begin{center}
\caption{A view of the axial power for this pin at the selected axial level.}
\label{fig:PWRPin2}
\includegraphics[width=\textwidth]{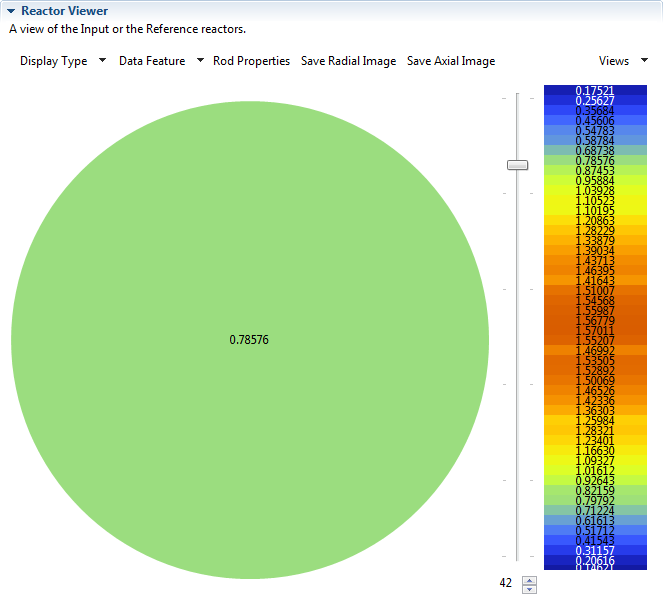}
\end{center}
\end{figure}

Figure\ \ref{fig:PWRPin1} shows the geometry of a fuel rod in problem 3a. The
rings of material are represented by the concentric circles of red, yellow, and
green areas for the fuel, fill gas, and clad, respectively. Similar to the view
for assemblies, the axial level can be adjusted using the slider, the spinner,
or the clickable axial view. The data view for this rod showing the pin power at
the 42nd axial level is shown in Fig.\ \ref{fig:PWRPin2}. The fuel is homogeneous
in problem 3a, so only one material block and ring are shown.

The views for SFR pins are not shown because they are, for the most part,
identical to the views for PWRs.

%
%
\subsection{Plots and Comparisons}

\begin{figure}[h!]
\begin{center}
\caption{A graph of the axial pin power across the B2, E4, and H7 pins in the
assembly for the ``input'' simulation performed with VERA.}
\label{fig:InputPlot}
\includegraphics[width=\textwidth]{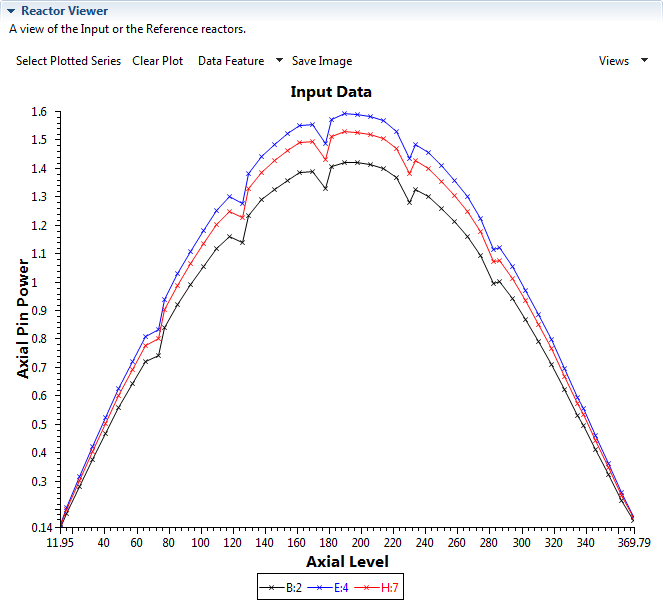}
\end{center}
\end{figure}

\begin{figure}[h!]
\begin{center}
\caption{A graph of the axial pin power across the B2, E4, and H7 pins in the
assembly for the ``reference'' simulation performed with KENO.}
\label{fig:RefPlot}
\includegraphics[width=\textwidth]{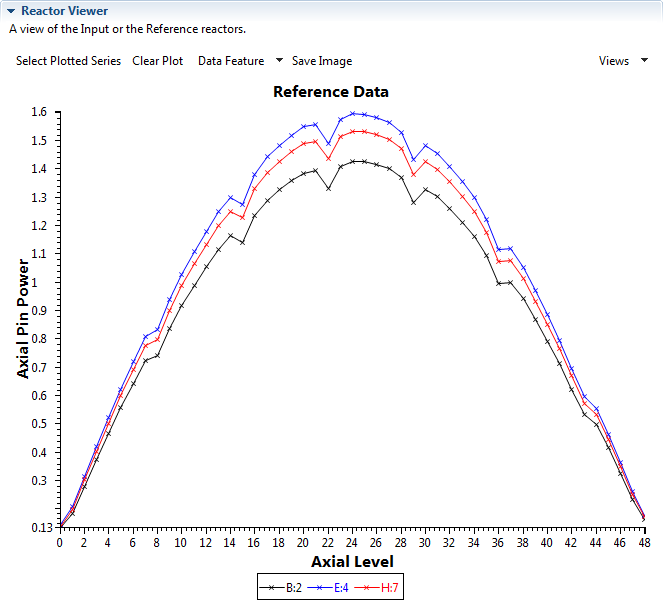}
\end{center}
\end{figure}

The user interface provides a limited plotting capability for information stored
on assemblies and pins/rods. This allows users to make quantitative comparisons
quickly and easily. Plots of the axial pin power for select fuel pins in problem
3a are shown in Figs.\ \ref{fig:InputPlot} and \ref{fig:RefPlot}. The height
from the bottom of the pin is shown on the horizontal axis, and the axial pin
power on the vertical axis.

%

\begin{figure}[h!]
\begin{center}
\caption{A graph showing the percentage difference in the axial pin powers for
the B2, E4, and H7 pins in the assembly.}
\label{fig:comparison}
\includegraphics[width=\textwidth]{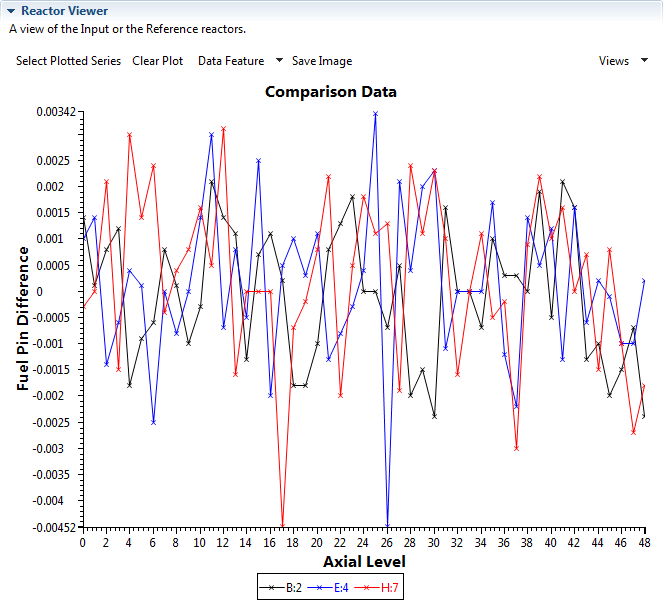}
\end{center}
\end{figure}

A plot showing the normalized percentage differences between the axial powers from
VERA and KENO for problem 3a is shown in Fig.\ \ref{fig:comparison}. This plot
was created using the external analysis routine described in
\ref{subsec:analysisTools} and shows very close agreement between the two
codes.

\section{Availability and Ongoing Work}

Source code and fully compiled binaries of this system are available as part of
NiCE at http://niceproject.sourceforge.net. The system is referred to as the
``Reactor Analyzer'' in NiCE, and several detailed tutorials are
available at the same website. Detailed source code documentation is also
available. Future versions of the system will be available in NiCE, but in the
near future that project will be ``converted'' into a new project at the Eclipse
Foundation called the ``Eclipse Integrated Computational Environment.'' The system
will then be available at http://www.eclipse.org. The current version is available
under a modified BSD license, and future releases will be available under
the Eclipse Public License.

Readers are encouraged to watch the recorded demonstrations published on
YouTube.com at http://www.youtube.com/jayjaybillings.

Detailed Unified Modeling Language models were developed as part of this work; they are also
available for download, although the format is proprietary. The authors will
gladly export the Unified Modeling Language models into a more friendly format upon request. 

The authors are open to feedback and contributions from readers. Those
interested in contributing to the work through testing, design improvements,
or source code development should contact the corresponding author.

\subsection{Limitations}

This system is relatively new and has several limitations. It remains
highly useful considering these limitations and will only improve
with time as they are addressed and other improvements added. 

Performance has not been tested with very large amounts of data and many time
steps. Scaling to a full core based on the 3a problem has performed well, after
work to optimize the code; but even at the size of a full core, problem 3a is not
very big. It is only resolved at 49 axial levels with two sets of state
point data and five distinct rod types.

The system has not been tested with real data from SFR simulations. Although the
authors do not expect that this will present any issues, that remains a
possibility.

The system needs to be modified to include more of the regular properties of
Parts. For example, the heights of rods and pins are determined by the total
heights of their material blocks, but it needs to be possible to retrieve this
value more easily. Plenum gas has not been considered.

\subsection{Planned Future Work}

There are many possible refinements to this system and a large amount of
upcoming work. The most immediate refinement will be the addition
of components to represent the pieces of a nuclear plant, such as generator,
pipes, and other ``plant-level elements.'' As previously mentioned, it
is also important to test the system for much larger amounts of data and with
the parallel I/O capabilities natively available in the HDF5 library.

Extensions to more languages are planned, including bindings for C, Fortran, and
Python. Each of these will be a wrapper around the C++ version, not a stand-alone
implementation.

The authors are currently working with members of the SHARP team \cite{SHARP}
to test the system with real data from simulations of SFRs. 

The model for pins and rods will be improved in the near future after tests with
Bison \cite{bison} later this year.

Extending the properties available in Parts is straightforward, and the authors
are working with collaborators to extend the set of properties for each Part.

\section{Conclusions}

Future simulations of new nuclear reactor designs will require new ways to
examine the results because of the high fidelity and resolution inherent in state-of-the-art
simulation codes. The system presented herein is capable of reducing the
analysis burden on both users and developers by organizing the results in an
intuitive, domain-specific way and providing easy-to-use I/O capabilities and a
user interface. Its application to a real-world problem with a 17 by 17
17 by 17 PWR fuel assembly from a VERA benchmark problem was shown by
generating plots of the percentage difference between axial powers from VERA
and KENO. Work remains to cover all of the different parts of LWRs and SFRs in
sufficient detail for widespread use, as well as to optimize the system for truly
large amounts of data in time and space.

Addressing the ``data problem'' will be critical to the success of the new
modeling and simulation capabilities in development. Certainly, it is necessary
to provide at least \textit{some} streamlined capability to examine the large
amounts of data coming from these simulations and, ideally, to make it possible
to discover interesting new physics in the results through data mining and
machine learning.

\section{Acknowledgments}
The authors are grateful to members of the NEAMS and CASL communities, others
whom we have interviewed, and our program managers. The authors are grateful for
the assistance of Greg Davidson, Jess Gehin, Andrew Godfrey, Ugur Mertyurek, and
John Turner from Oak Ridge National Laboratory (ORNL) and Justin Thomas from Argonne
National Laboratory. The authors are also grateful for the financial support
provided by the NEAMS, Advanced Reactor Concepts, and CASL programs. 

The authors are especially thankful to David Pointer of ORNL
for his assistance, guidance, and continued support during the
performance of this work and the development of the paper.

This work has been supported by the US Department of Energy, Office of
Nuclear Energy, and by the ORNL Postgraduate Research Participation Program, which
is sponsored by ORNL and administered jointly by ORNL and the Oak Ridge
Institute for Science and Education (ORISE). ORNL is managed by UT-Battelle, LLC,
for the US Department of Energy under contract no. DE-AC05-00OR22725. ORISE
is managed by Oak Ridge Associated Universities for the US Department of
Energy under contract no. DE-AC05-00OR22750.

\section{References}
\bibliography{references}
\bibliographystyle{plain}
\end{document}